\documentstyle[aps,pre,draft,graphicx,floats]{revtex}
\begin{document}
 \baselineskip=24pt
 \draft
%{\wideabs{ 
\title{Helix vs.~Sheet Formation in a Small Peptide}   
%\title{Environment and Secondary Structure Formation in a small Peptide}   
\author{Yong Peng and
        Ulrich H.E. Hansmann \footnote{ito whom all correspondence
                       should be addressed. E-mail: hansmann@mtu.edu}}
\address{Department of Physics, Michigan Technological University,
         Houghton, MI 49931-1291, USA}
\date{\today}
\maketitle

\begin{abstract}
Segments with the amino acid sequence EKAYLRT appear 
in natural occurring proteins  both in  
$\alpha$-helices and $\beta$-sheets. For this reason, we
have use this peptide to study how  secondary
structure formation in proteins depends on the 
local environment. Our data rely on multicanonical Monte Carlo 
simulations where the interactions among all atoms are taken 
into account. Results in gas phase are compared with that in
an implicit solvent. We find that both in gas phase and solvated
EKAYLRT forms an $\alpha$-helix when not interacting with other molecules.
However, in the vicinity of a $\beta$-strand, the peptide 
forms  a $\beta$-strand. Because of this change in secondary structure
our  peptide  may provide a
simple model for the $\alpha \rightarrow \beta$ transition 
that is supposedly  related to the outbreak
of Prion diseases and similar illnesses.
\end{abstract}
%\pacs{}
%}}
%\begin{multicols}{2}

%%%%%%%%%%%%%%%%%%%%%%%%%%%%%%%%%%%%%%%%%%%%%%%%%%%%%%%%%%%%%%%%%
\section{Introduction}
Despite considerable progress over the last decade the problem
of predicting the biological active structure of a protein solely
from the sequence of amino acids has remained a formidable problem.
More successful have been attempts to predict only the secondary
structure. Given the protein sequence it is  today possible to
determine the distribution and location of $\alpha$-helices and 
$\beta$-sheets with up to $90$\% probability. This
high success rate indicates a close relation between sequence 
information and  secondary structure. 
However, two observations indicate that this relation is not 
a simple one.  First, certain sequences can form 
either $\alpha$-helices or $\beta$-sheets \cite{Su}. The most
prominent example is the 11-residue Chameleon peptide \cite{MiKi}
that folds as an $\alpha-$helix when replacing residue 22-32 
of the primary sequence of the IgG-binding domain of protein G 
(57 amino acids), but as a $\beta$-strand when inserted instead
of residues 42-52. Secondly, it has become clear over the last
years that miss-folding of proteins, often involving formation
of $\beta$-sheets instead of $\alpha$-helices, and subsequent
aggregation is the cause of various illnesses including 
Alzheimer's disease, BSE and other Prion diseases. Hence, 
it is important to understand in detail how  secondary structure
formation and its role in the folding process depends on 
the intrinsic properties of the protein and the interaction with
the surrounding environment.

In order to study these questions we have simulated a peptide
whose sequence of amino acids  EKAYLRT 
(glutamine - lysine - alanine - tyrosine - leucine - arginine - threonine)
appears in natural occurring proteins with significant frequency 
at positions of both $\alpha$-helices and $\beta$-sheets. Our
present work differs therefore from previous investigations where
we have focused on helix-formation and folding in homopolymers and 
artificial peptides \cite{OH95b,HO98c,AH99b,PH01g,PHA02,AH01i}. Unlike
these molecules that have a strong intrinsic tendency
to form one specific kind of secondary structure elements 
($\alpha$-helices), EKAYLRT allows one to research   
the selection of either helix or sheet, or the transition between
these two secondary structures, as a function of external
factors.

Our work differs from  similar approaches \cite{Chen2,Chen1}
in that we study not minimal models but simulate detailed 
representations of our peptides where the interactions 
between all atoms are taken 
into account. EKAYLRT is simulated both in gas phase and with an implicit 
solvent.  Quantities such as energy, specific heat, sheetness and 
helicity are calculated as functions of temperature.  
We find that both the solvated molecule and  EKAYLRT in gas phase  form
an $\alpha$-helix when not interacting with other molecules.
However, in the vicinity of a $\beta$-sheet the peptide prefers
also to form strand. Because of the resulting ``auto-catalytic''
property our  peptide may therefore  provide a simple model for 
the $\alpha \rightarrow \beta$ transition and the resulting 
aggregation process in some proteins that supposedly is related to 
the outbreak to neurological diseases such as Alzheimer's and the
Prion diseases.

%%%%%%%%%%%%%%%%%%%%%%%%%%%%%%%%%%%%%%%%%%%%%%%%%%%%%%%%%%%%%%%%%%%%%%%%

\section{Methods}
Our aim here is to research  how  secondary structure formation and its 
role in the folding process depend on either the intrinsic properties of 
a protein or its interaction with the surrounding environment.
For this purpose, we have considered  detailed, all-atom representations 
of peptides that are based on the sequence of amino acids EKAYLRT.  
 To be more specific, the peptide NH$_2$-EKAYLRT-COOH 
is studied both as an isolated molecule and interacting with another EKAYLRT
peptide that is held in a $\beta$-strand conformation. Since our
program package SMMP \cite{SMMP} in its current version allows
only the simulation of single peptides we have modeled the latter
case by considering the peptide NH$_2$-EKAYLRT-GGGG-EKAYLRT-COOH,
with the C-terminal EKAYLRT residues kept as a  $\beta$-strand. The
four  glycine residues form a flexible chain that hold the two
peptides together but allows their relative positions to vary. 
The underlying assumption is that the interaction between the two
EKAYLRT chains is the dominant term and their interaction with the
glycine residues can be neglected.

The intra-molecular interactions are described by a 
standard force field, ECEPP/3,\cite{EC3}  (as implemented in the  
program package SMMP \cite{SMMP}) and  are given by:
\begin{eqnarray}
E_{ECEPP/3} & = & E_{C} + E_{vdW} + E_{HB} + E_{tor},\\
E_{C}  & = & \sum_{(i,j)} \frac{332q_i q_j}{\epsilon\, r_{ij}},\\
E_{vdW} & = & \sum_{(i,j)} \left( \frac{A_{ij}}{r^{12}_{ij}}
                                - \frac{B_{ij}}{r^6_{ij}} \right),\\
E_{HB}  & = & \sum_{(i,j)} \left( \frac{C_{ij}}{r^{12}_{ij}}
                                - \frac{D_{ij}}{r^{10}_{ij}} \right),\\
E_{tor}& = & \sum_l U_l \left( 1 \pm \cos (n_l \chi_l ) \right).
\end{eqnarray}
Here, $r_{ij}$ (in \AA) is the distance between the atoms $i$ and $j$,
 and $\chi_l$ is the $l$-th torsion angle. The peptide bond angles 
are set to their common value $\omega = 180^{\circ}$. We further 
assume for the  electrostatic permittivity in the  protein 
interior  $\varepsilon=2$ (its common value in ECEPP simulations). 

Simulations of our peptide EKAYLRT in gas phase are compared with
such where the interaction of the peptide with surrounding water
is approximated by an implicit solvent \cite{oons}: 
\begin{equation}
E = E_{ECEPP/3} + E_{solv} \quad {\rm with} \quad 
                  E_{solv} = \sum_i \sigma_i A_i~.
\label{solvent}
\end{equation}
Here, $E_{solv}$ is the solvation energy and thought to be
proportional to the solvent accessible surface area $A_i$ of 
the $i$th atom.  The parameters $\sigma_i$ are experimentally determined
proportionality factors. 

Simulations of such detailed protein models are extremely difficult. 
This is because the various competing interactions lead to
multitude of local energy minima separated by high  barriers.  Hence, 
in the low-temperature region, canonical Monte Carlo or molecular 
dynamics simulations will  get trapped in one of these minima and 
not thermalize within the available CPU time. Only with 
the introduction of new and sophisticated algorithms such as
{\it generalized-ensemble} techniques \cite{MyReview}, is it
possible to alleviate this problem in  protein simulations \cite{HO}.
For this reason,  our investigations rely on the use of one of
these techniques, multicanonical sampling \cite{MU},  where
conformations with energy $E$ are assigned a weight
$  w_{mu} (E)\propto 1/n(E)$ ($n(E)$ is that is  the density of 
states).  A  simulation with this weight will generate a 1D 
random walk in the energy space and lead to a uniform distribution 
of energy:
\begin{equation}
  P_{mu}(E) \,  \propto \,  n(E)~w_{mu}(E) = {\rm const}~.
\label{eqmu}
\end{equation}
Since a large range of energies are sampled, one can
use the reweighting techniques \cite{FS} to  calculate thermodynamic
quantities over a wide range of temperatures $T$ by
\begin{equation}
<{\cal{A}}>_T ~=~ \frac{{\int dx~{\cal{A}}(x)~w^{-1}(E(x))~
                 e^{-\beta E(x)}}}
              {{\int dx~w^{-1}(E(x))~e^{-\beta E(x)}}}~,
\label{eqrw}
\end{equation}
where $x$ stands for configurations and $\beta=1/k_BT$ is the 
inverse temperature.  Estimators for the multicanonical weights 
$w(E) = n^{-1}(E) = \exp(-S(E))$ can be  calculated with the
iterative procedures described in Refs.~\cite{PH01g}. 

In our case we  needed between 100,000 and 200,000  sweeps for 
the  weight factor calculations.  All thermodynamic quantities are  
then estimated from one production run of $2,000,000$  Monte Carlo sweeps 
that followed $10,000$ sweeps for ``thermalization''.  Our simulations  
start from  completely random initial conformations (Hot Start) 
and  one Monte Carlo sweep updates every torsion angle of the peptide 
once.  At the end of every 4th  sweep, we store the total
energy $E_{Tot}$, the ECEPP/3 energy $E_{ECEPP/3}$, its partial terms
$E_C, E_{LJ}, E_{HB}$ and $E_{tor}$,  the solvation energy
$E_{Solv}$,  the corresponding end-to-end distance $d_{e-e}$, 
and  the number $n_H$ ($n_B$) of helical (sheet) residues. 
Here, we follow  previous work \cite{OH95b} 
and consider a residue as helical if its backbone angle $(\phi,\psi)$ are 
within the range $(-70^{\circ}\pm 30^{\circ},-37^{\circ}\pm30^{\circ})$.
Similar, a residue is assumed to be ``sheet-like'' if $(\phi,\psi)$ are
within the range $(-140^{\circ}\pm 40^{\circ},140^{\circ}\pm40^{\circ})$.

%%%%%%%%%%%%%%%%%%%%%%%%%%%%%%%%%%%%%%%%%%%%%%%%%%%%%%%%%%%%%%%%%%%%%%%%
\section{Results and Discussion}
We start with presenting our results for a single EKAYLRT molecule
that is not interacting with other molecules. We display for this
peptide in Fig.~1 the average helicity $<n_H>(T)$ as a function of 
temperature.  Shown are data  obtained in gas-phase (GP) and such for 
simulations  that rely on a solvent accessible surface area term (ASA)
to approximate protein-water interactions.  We observe 
in both cases a steep helix-coil
transition that separates a high-temperature region with little
helicity from a low-temperature region where most of the residues
are part of an $\alpha$-helix. The location of this transition
can be determined from the corresponding peaks in the specific heat
$C(T)$ that are drawn in the inlet. We find as the helix-coil
transition temperature of EKAYLRT in gas phase  
$T^{GP}_{hc} = 445\pm15$ K.  The more pronounced peak for the 
solvated molecule indicates  a temperature $T_{hc}^{ASA} = 340\pm10$ K 
that is considerably lower than the one in 
gas phase. Unphysiologically high helix-coil transition temperatures
in gas phase, and their shift toward a more sensible temperature range 
when an implicit solvent is introduced, have been  also
observed in our earlier work on homopolymers \cite{PH01g,PHA02}. 

We show in Fig.~2 as an example for the helical configurations that 
dominate below  $T_{HC}$   the lowest energy configuration found
in a simulation of the solvated peptide ($E_{Tot} = -69.6$ kcal/mol).
The lowest energy configuration in gas phase ($E_{Tot}=E_{ECEPP/3}= -28.0$
kcal/mol) is also a helix (structure not shown).  The energy of these
helical structures is by $\approx 25$ kcal/mol lower than the ones of the
lowest found ``sheet-like'' configurations: $E_{Tot} = -43.8$ kcal/mol
for the solvated peptide and $E_{Tot} = E_{ECEPP/3} = -3.1$ kcal/mol
for EKAYLRT  in gas phase.

The preference for helical structures can be also seen in Fig.~3a 
where we display the free energy $\Delta G$ at $T=300$ K 
as a function of helicity $n_H$  and ``sheetness'' $n_B$.  Note that 
for convenience we have chosen a normalization where the minimum 
in free energy takes a value
of zero.  Both in gas phase and for the solvated 
molecule a funnel-like free-energy landscape is formed, with the
free energy minimum at $n_H = 5$, i.e. for maximal helicity 
(since the two terminal ends are flexible and will usually not 
be part of an helix, a fully formed helix has a length $n_H=5$ instead
of $n_H=7$). The absolute value of the free energy 
difference between coil and helix is much larger for the peptide 
in gas phase ($\Delta G \approx -5$ kcal/mol) than it is for the 
solvated molecule ($\Delta G \approx -2$ kcal/mol) indicating 
that the helix-coil transition is stronger for EKAYLRT in gas
phase than for the molecule in an implicit solvent. This is in agreement with 
earlier work where we have found similar results for polyalanine 
chains \cite{PHA02}.  The corresponding projection of the free
energy landscape on the ``sheetness'' $n_B$ in Fig.~3b shows the
opposite picture; the free energy increase with the number of residues
whose backbone dihedral angles takes values that are common in a
$beta$-sheet. Coil structures are at $T=300$ K favored over 
sheet-like structures by $\Delta G \approx 5$ kcal/mol in the 
implicit solvent and by $\Delta G \approx 8$ kcal/mol in gas phase.

The observed   form of the free-energy landscape is caused  solely by the
intra-molecular interactions. This can be seen in Fig.~4 where
we plot for solvated EKAYLRT the total energy $E_{Tot}$, the
internal energy $E_{ECEPP/3}$ and the solvation energy $E_{Solv}$
as a function of temperature. Here, we have normalized all energy
terms in such way that their value at $n_H = m0$ ($n_B=0$) is zero.
Both $E_{Tot}$ and $E_{ECEPP/3}$ decrease
with growing number of residues that are part of a helix while
$E_{Solv}$ increases (Fig.~4a). Hence, the protein-water interaction
term opposes helix-formation. This result is reasonable as
the protein-water hydrogen bonds  compete with the intra-molecular 
hydrogen bonding in an $\alpha$-helix and therefore
weaken helix-formation in solution. However, the loss in solvation
energy of $\Delta \approx 4$ kcal/mol with helix-formation is small
when compared with the gain in $E_{ECEPP/3} \approx -16$ kcal/mol,
and on average, a completely formed helix ($n_H =5$) has a total
energy that is by $\Delta E_{Tot} \approx  -12$ kcal/mol lower
than a coil configuration ($n_H=0$). Not surprisingly, we observe
the opposite behavior in Fig.~4b where we plot the same three energies
as a function of ``sheetness'' $n_B$. Sheet-like configurations with
large numbers $n_B$ have higher internal energy $E_{ECEPP/3}$ than
such with $n_B =0$ while the solvation energy $E_{solv}$ is lower.

Hence,  while at $T=300$ K the protein-water interaction seems 
to favor strands and opposes helix-formation, the physics of 
our molecule is dominated by the intra-molecular energies that
lead to a strong preference for $\alpha$-helix formation. Fig.~5
indicates that this behavior is mainly due to the van der Waals
interaction between the atoms in the peptide.  In this figure,
we display as a function of temperature  besides the van der 
Waals term $<E_{vdW}>$ also the
other partial energies that together make up $E_{ECEPP/3})$:
the average electrostatic energy  $<E_C>$, the hydrogen-bond 
energy $<E_{HB}>$ and the  torsion energy $<E_{Tor}>$. 

Our results so far indicate that the peptide EKAYLRT has a  
intrinsic tendency to form helices. Strands have of order 
$\approx 30$ kcal/mol higher free energies and are rarely
observed. This result is independent on whether the 
molecule is in gas phase or simulated with an implicit solvent.
However, EKAYLRT  appears {\it within}
proteins  both in   helices and $\beta$-sheets. It follows that 
sheet formation has to be due to the interaction of the peptide
with its surrounding. We conjecture that  EKAYLRT forms a $\beta$-sheet
if it is in the proximity of another strand. Especially,
we assume that this process also happens if the peptide is 
close to another EKAYLRT peptide that is already in a strand
configuration. Unfortunately, the 
present version of SMMP does not allow the simulation of 
two interacting  proteins. Hence, in order to test our conjecture, 
we have studied instead the peptide NH$_2$-EKAYLRT-GGGG-EKAYLRT-COOH
with the C-terminal EKAYLRT residues kept  as $\beta$-strand. The
four  glycine residues form a flexible chain that hold the two
EKAYLRT-units together but allows their relative positions to vary. 
We refer to the so constructed peptide as molecule `A`.
 
The end-to-end distance $d_{e-e}$ is a measure for the separation of
the two EKAYLRT chains. Our conjecture implies that for large 
values of $d_{e-e}$ the N-terminal  EKAYLRT assumes an $\alpha$-helix
while for small values of $d_{e-e}$ (i.e. close proximity to
the C-terminal  EKAYLRT that forms a strand) it should
assume a $\beta$-sheet-configuration. We have therefore 
calculated from the multicanonical simulation of molecule `A`
the helicity and sheetness of the  N-Terminal EKAYLRT at $T=300 K$.
Both quantities are displayed in Fig.~6.  Two regions are observed.
For $d_{e-e}> \approx 16$ \AA\ the N-terminal EKAYLRT chain forms
a complete helix and strands are rarely observed. Hence,
for these distances the N-terminal chain has a similar behavior 
as the isolated EKAYLRT-peptide.
However, for decreasing end-to-end distance, the helicity also decreases
and vanishes for $d_{e-e} < \approx 10$ \AA.  At the same time,
the sheetness increases and the peptide forms a $\beta$-sheet for
$d_{e-e} \approx 5-6$ \AA. Note that the average potential energy of helical 
configurations is with $<E_{Tot}> =  -24.9 (1.6)$ kcal/mol within
the errorbars equal to that  of  sheet-like
configurations ($<E_{Tot}> = -23.4(2.9)$ kcal/mol).

In Fig.~7,  the projection of the free-energy landscape  at room 
temperature ($T=300$ K) on the helicity and sheetness of the 
N-terminal EKAYLRT residues is drawn. For convenience, we have set 
in this figure the
lowest-found value of the free-energy to zero as energies are only
defined up to an additive constant. The contour lines are spaced by
2 kcal/mol. The free-energy landscape is only plotted for values
of $G \le 25$ kcal/mol  as values of the free energy grow rapidly outside
of the drawn area. We observe again two minima, corresponding 
to fully formed helix and $\beta$-strands. Examples of configurations that 
correspond  to the two minima are shown in Fig.~8. Both minima have 
comparable free energies and are separated by  barriers of only 2 kcal/mol 
allowing an easy interchange between the two forms.

In order to understand in more detail why EKAYLRT forms a $\beta$-strand 
when close to a molecule that is already in a $\beta$-sheet
form, we have performed further simulations of 
 NH$_2$-EKAYLRT-GGGG-EKAYLRT-COOH holding now not only
the C-terminal EKAYLRT-residues as a $\beta$-strand  but forcing 
also the four connecting glycine residues into a turn. We refer to 
the so defined peptide as molecule `B'.  The N-terminal EKAYLRT-residues  
are now by construction in close proximity to
the C-terminal EKAYLRT-strand.  Hence, we expect that  at room
temperature the N-terminal EKAYLRT chain will also form a
$\beta$-strand. This conjecture is supported by Fig.~9 where we plot
the average ``sheetness'' $n_B$ of the  N-terminal EKAYLRT-residues
as a function of temperature. Both in gas-phase and for simulations
with a solvent accessible surface term,
we find that on average more than 5 of the 7 residues
are part of a sheet-like structure. $<n_B>$ decreases smoothly with
growing temperature and the maximum in the specific heat is shallow.
The transition is more pronounced for the peptide in an implicit solvent
than for the one in gas-phase, and shifted toward lower temperatures. 

Unlike our previous simulations where the glycine residues could freely
move,  a large percentage of configurations are now at room temperature in 
a sheet-form. The
increased statistics of these configurations allows for a better analysis
of the factors that help to overcome the intrinsic propensity of
EKAYLRT  to form an $\alpha$-helix and lead to 
a $\beta$-sheet. Table 1 lists the  differences of various energies 
between structures where the N-terminal EKAYLRT unit is a $\beta$-strand
with structures where these residues form an $\alpha$-helix.
Values are listed for the whole molecule `B' and such restricted to
the  N-terminal EKAYLRT chains. Also listed are 
the differences of both terms. The latter quantity is a measure 
for the interactions between these seven residues and the rest of the 
molecule (that is kept fixed).
       
We see from this table that at $T=300$ K configurations  with the
N-terminal EKAYLRT chain  in a sheet are energetically favored 
by 7 kcal/mol over such where these residues form an 
$\alpha$-helix. This energy-bias is found
for all partial energies with the exception of the solvation
energy term $E_{Solv}$ and the torsion energy term $E_{Tor}$.
While their values seems to indicate a slight preference for
helical  over sheet-like configurations, they are  
within the errorbars compatible with zero   suggesting that
both terms show no preference for one of the two forms.
 
Because of the  bias in the internal energies $E^{Molecule~`B'}$
of the whole molecule `B', $\beta$-sheet-like conformations of the
N-terminal EKAYLRT chain dominate at room temperature. However,
the  behavior of the various energy terms is different when one 
considers only the contributions by these seven residues.  
With the exception of the solvation energy $E^{EKAYLRT}_{Solv}$, 
that favors $\beta$-strands, all energy terms favor now a helix. On average,
helical structures have at $T=300$ K a $14.5(1.4)$ kcal/mol lower energy
$E^{EKAYLRT}$ than strands when only the interaction between atoms in
these seven residues are considered. Hence, their behavior
is qualitatively the same as for the   isolated EKAYLRT-peptide
where we also observed a strong bias toward helical conformations.
Again, we find also that the van der Waals energy $E_{vdW}$ 
is the dominant term. It follows
that the $\beta$-sheet configurations that dominate when the
EKAYLRT residues are  build into molecule `B' are caused by
 the interaction between this chain and   the ``background'' of the 
rest of the molecule. Since
energies are additive, we can calculate this ``background''-field by
\begin{equation}
E^{Background} = E^{Molecule `B'} - E^{EKAYLRT}~.
\label{background}
\end{equation}
The strength of the interaction between the peptide and the 
``background field'' of the rest of molecule `B' can be seen from
the large energy-difference of 
$\Delta E^{Background}_{Tot} = -21.8 (1.6)$ kcal/mol
by that these interactions favor a strand. 
The main contribution comes from the
van der Waals term ($\Delta E^{Background}_{vdW} = 15.7 (1.1)$ kcal/mol)
which is almost three times as large as the electrostatic and
torsion energy terms. Note  that Eq.~\ref{background} tell us also
that the ``background''  given by the fixed parts of molecule`B' 
raises the solvation energy difference 
$\Delta E^{EKAYLRT}_{Solv}= -3.1(2)$ kcal/mol of the EKAYLRT-chain by
$\Delta E^{Background}_{Solv} = 3.7(3)$ kcal/mol to a value
of $\Delta E^{Molecule `B'}_{Solv} = 0.6(3)$ kcal/mol for the whole
system. This is because  the term $\Delta E^{EKAYLRT}_{Solv}$ is due
to the competition between  hydrogen-bond formation in an $\alpha$-helix
and hydrogen-bond formation between the peptide and the surrounding water.
However, in Molecule `B' the peptide is geometrically constraint in such
a way that this competition is replaced by one between  hydrogen-bond 
formation in an $\alpha$-helix of the N-terminal EKAYLRT on one side, and
formations of hydrogen-bonds between the peptide and the C-terminal
EKAYLRT-residues on the other side (see also the opposite sign in the terms 
$\Delta E^{EKAYLRT}_{HB}$ and $\Delta E^{Background}_{HB}$ in table 1).
As a result, both the solvation energy difference 
$\Delta E^{Molecule~`B'}_{Solv}$ and hydrogen-bond energy difference
$\Delta E^{Molecule~`B'}_{HB}$ are marginal. Instead, the preference
of $\beta$-sheet configurations for  EKAYLRT in the ``background''
of the fixed rest of Molecule `B' seems to be mainly due to the
electrostatic and van der Waal's energies. This is reasonable: 
a $\beta$-sheet
conformation allows for an average closer distance between the
atoms of the N-terminal EKAYLRT-chain and the existing $\beta$-strand of
the C-terminal EKAYLRT-residues, decreasing in this way the
van der Waals energy. At the same time, the alignment of the two
$\beta$-strands leads also to a favorable alignment of the dipole moments
associated with each residue lowering therefore the electrostatic energy.
 We conjecture that without the stereometric constraints
imposed by the connecting glycine residues the two strands
would move together and aggregate as the energy gain increases with
decreasing distance between them.

Our above presented results for Molecule `A' and Molecule `B' suggest
auto-catalytic properties for EKAYLRT: if the peptide forms a strand,
in becomes energetically favorable for other nearby EKAYLRT molecules
to transform themselves into a sheet (instead of the normally preferred
helix), and eventually to aggregate with the first one. This 
behavior is similar to the mechanism thought to be responsible 
for the outbreak of neuro-degenerative  illnesses such as Alzheimer's or
the Prion diseases. Outbreak of theses illnesses is 
 associated with the appearance of a miss-folded 
structure  that differs from the correctly
folded one by a $\beta$-sheet instead of an $\alpha$-helix. The miss-folded
structure is thought to be auto-catalytic, that is its presence leads
to a structural transition by which the correctly folded (helical)
structure changes into the harmful $\beta$-sheet form.  Hence, 
peptides that contain the sequence of amino acids EKAYLRT 
can serve as simple  models  to study 
these $\alpha \rightarrow \beta$-transition and the mechanism of 
Prion diseases. For instance, our investigation suggest that the
formation of $\beta$-sheets can be minimized by shielding the surface
area of already existing $\beta$-sheet forms minimizing in this way
the van der Waals interaction. Another possibility may be to 
introduce metal ions that alter the electrostatic interaction
decreasing in this way the energy bias toward $\beta$-sheets.

%%%%%%%%%%%%%%%%%%%%%%%%%%%%%%%%%%%%%%%%%%%%%%%%%%%%%%%%%%%%%%%%%%%%%%%%

\section{Conclusion}
We have performed multicanonical simulations of peptides that contain
the sequence of amino acids  EKAYLRT. We  find that the  
EKAYLRT-peptide itself has both in gas-phase and in solution 
an intrinsic tendency to form  an $\alpha$-helix.  However, the
peptide assumes a $\beta$-sheet form when close to another strand.
The transition from an $\alpha$-helix toward a $\beta$-sheet is caused 
by strong van der Waals und electrostatic energy terms that favor
the $\beta$-sheet form when  EKAYLRT is in close proximity to another
strand. This ``auto-catalytic''
property of  EKAYLRT, that induces strand-formation in other
EKAYLRT molecules when in a $\beta$-sheet configuration, suggests
that the EKAYLRT based peptides can serve as a simple model for 
the $\alpha \rightarrow \beta$-transitions and successive aggregation
that are supposed to be related to the outbreak of various illnesses
such as Alzheimer's or the Prion diseases.\\

%%%%%%%%%%%%%%%%%%%%%%%%%%%%%%%%%%%%%%%%%%%%%%%%%%%%%%%%%%%%%%%%%%%%%%%%

\noindent
{\bf Acknowledgments}: \\
 U.H.  gratefully acknowledges support by a research grant
from the National Science Foundation (CHE-9981874).
Part of this article  was written while U.H. visited  the University 
of Central Florida in Orlando, FL.  He thanks  the UCF Physics Department 
for kind hospitality. 

%\end{multicols}

%%%%%%%%%%%%%%%%%%%%%%%%%%%%%%%%%%%%%%%%%%%%%%%%%%%%%%%%%%%%%%%%%%%%%%%%

\vfil

%%%%%%%%%%%%%%%%%%%%%%%%%%%%%%%%%%%%%%%%%%%%%%%%%%%%%%%%%%%%%%%%%%%%%%%%
\clearpage
\newpage
\noindent {\huge Tables:}\\

\noindent
\begin{table}[h]
\begin{tabular}{c|c|c|c}
%\hline
      & Molecule `B'  &N-terminal EKAYLRT residues only    & Background\cr
\hline
$\Delta E_{Tot}$  & -7.2(9)   & 14.5(1.4)      & -21.8(1.6)\\
$\Delta E_{Solv}$ &  0.6(3)   & -3.1(2)        &   3.7(3)\\
$\Delta E_{EL}$   & -4.3(3)   &  0.6(1)        &  -4.9(3)\\
$\Delta E_{vdW}$  & -3.2(8)   & 12.5(9)        & -15.7(1.1)\\
$\Delta E_{HB}$   & -0.8(2)   &  4.0(2)        &  -4.8(3)\\
$\Delta E_{Tor}$   &  0.5(4)   &  0.5(4)        &   0.0\\
%\hline
\end{tabular}
\caption{Energy differences between ``sheet'' and helix configurations
         at room temperature for various energy terms as calculated from
         a multicanonical simulation of molecule `B'.}
\end{table} 
%%%%%%%%%%%%%%%%%%%%%%%%%%%%%%%%%%%%%%%%%%%%%%%%%%%%%%%%%%%%%%%%%%%%%%%%
\newpage
\noindent {\huge Figure Captions:}\\
\begin{description}
\item [Fig.~1] The average number $<n_H>$ of helical residues as
               a function of temperature $T$ for EKAYLRT in gas phase
               (GP) and simulated with an implicit solvent term (ASA). The
               specific heat $C(T)$ as function of temperature $T$ 
               is displayed in the inlet. All results rely on 
               multicanonical simulations of 2,000,000 sweeps each.
\item [Fig.~2] Lowest energy configuration of EKAYLRT as found in
               a multicanonical simulation of 2,000,000 sweeps
               using an implicit solvent to approximate the peptide-water
               interactions.
\item [Fig.~3] The free energy $\Delta G$ at $T=300$ K as a function
               of ({\it bottom}) helicity $n_H$ and ({\it top}) sheetness $n_B$
               for EKAYLRT in gas phase ($\Box$) and simulated with an implicit 
               solvent term ($\Delta$). The free energy is
               normalized in such a way that its minimum value is set
               to zero. All results are calculated from a multicanonical
               simulation of 2,000,000 sweeps.
\item [Fig.~4] The average total energy $<E_{Tot}>$ ($\Box$), intra-molecular
               energy $<E_{ECEPP/3}>$ ($\Delta$) and solvation energy 
               $<E_{Solv}>$ ($\circ)$ of
               EKAYLRT at $T=300$ K as a function of ({\it bottom}) 
               helicity $n_H$
               and ({\it top}) sheetness $n_B$. All energies are normalized in 
               such way that their value at $n_H=0$ ($n_B=0$) is zero.
               All results are calculated from a multicanonical
               simulation of 2,000,000 sweeps using an implicit
               solvent model to approximate peptide-water interactions.
\item [Fig.~5] The average partial energies $<E_C>$, $E_{vdW}$,$E_{HB}$
               and $E_{Tor}$ that together make up the ECEPP/3 energy
               $E_{ECEPP/3}$ as a function of temperature $T$. All
               terms are normalized in such way that their value for
               $T=1000$ K is zero. All results are calculated from a 
               multicanonical simulation of 2,000,000 sweeps using 
               an implicit solvent model to approximate peptide-water 
               interactions.
\item [Fig.~6] The average helicity $<n_H>$ and sheetness $<n_B>$ at
               $T=300$ K of the  N-Terminal EKAYLRT residues as a
               function of the end-to-end distance $d_{e-e}$. All results
               are calculated from a multicanonical simulation of 
               2,000,000 sweeps. 
\item [Fig.~7] The free-energy landscape  of molecule `A'
               at room temperature ($T=300$ K) as a function of 
               helicity $<n_H>$ and sheetness $<n_B>$. The global minimum
               is set to zero and the contour lines are spaced by 2 kcal/mol.
\item [Fig.~8] Low-energy configurations of molecule `A'  
               as found in a multicanonical
               simulation of 2,000,000 sweeps. The one in (a) is the
               lowest-energy configuration where the N-terminal 
               EKAYLRT-residues  form an $\alpha$-helix; the one
               in (b) where they form  a $\beta$-sheet.
\item [Fig.~9] The average ``sheetness'' $<n_B>$ of the 
               N-terminal EKAYLRT-residues of molecule `B' 
               as a function of temperature. 
\end{description}

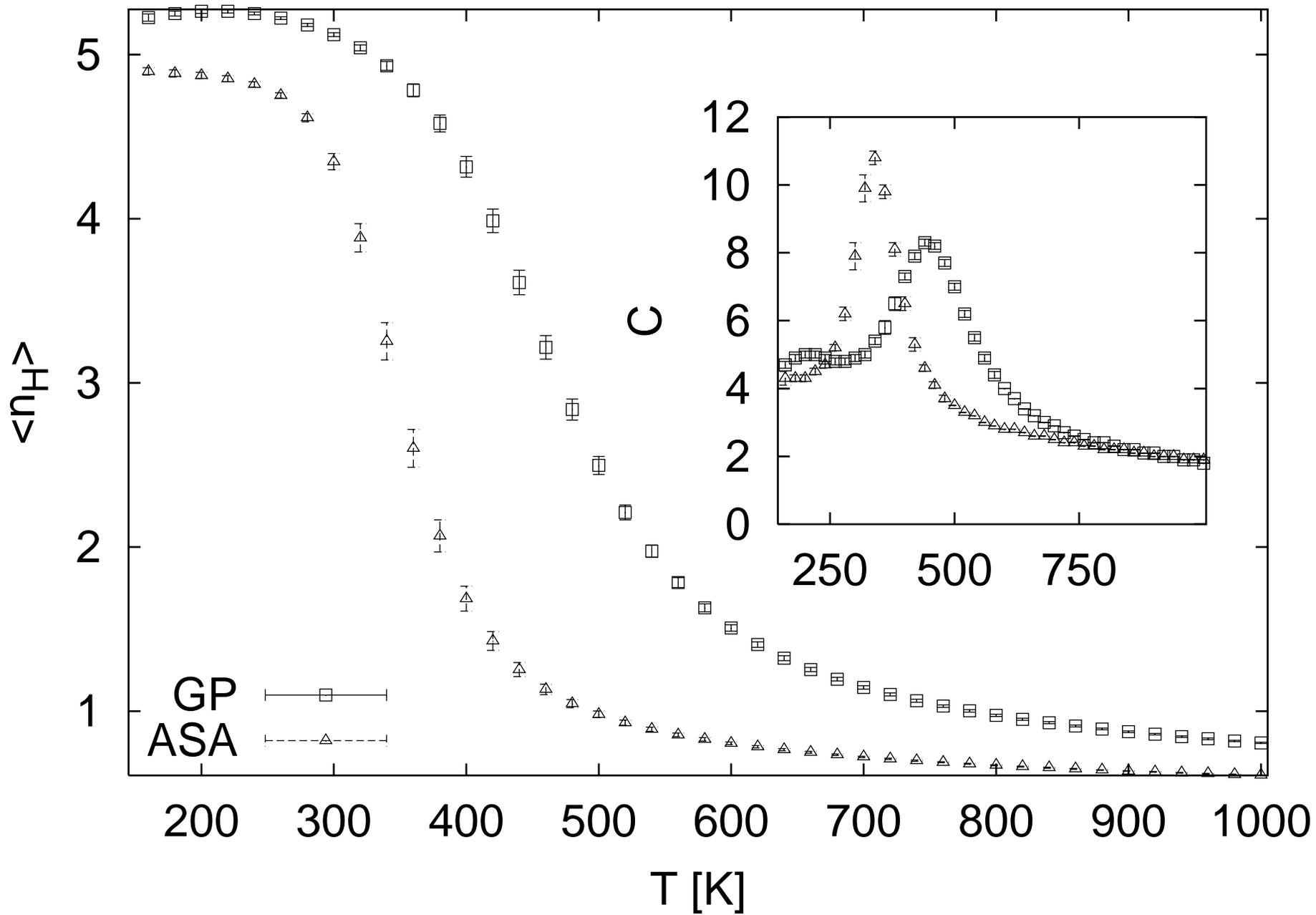
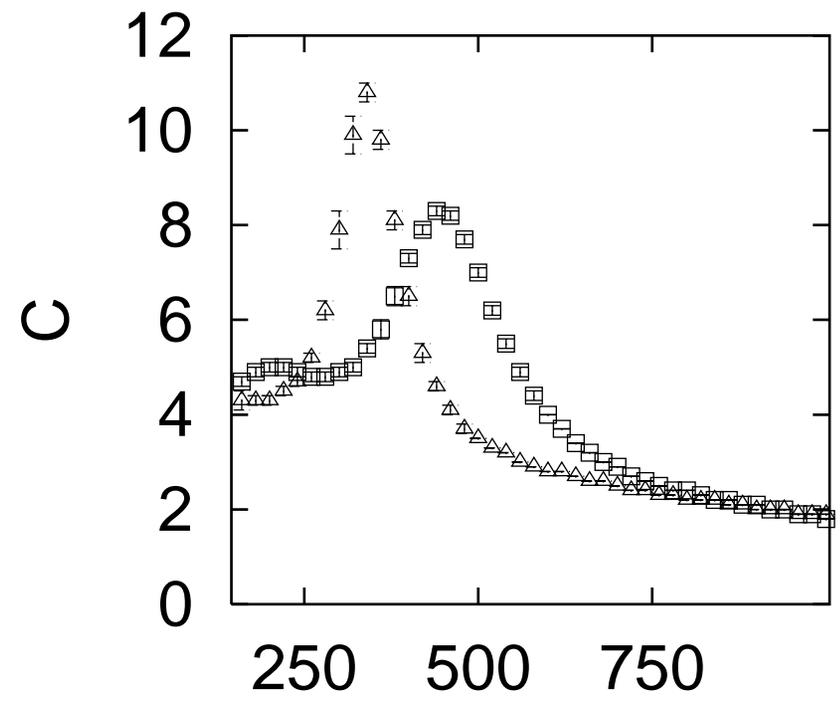
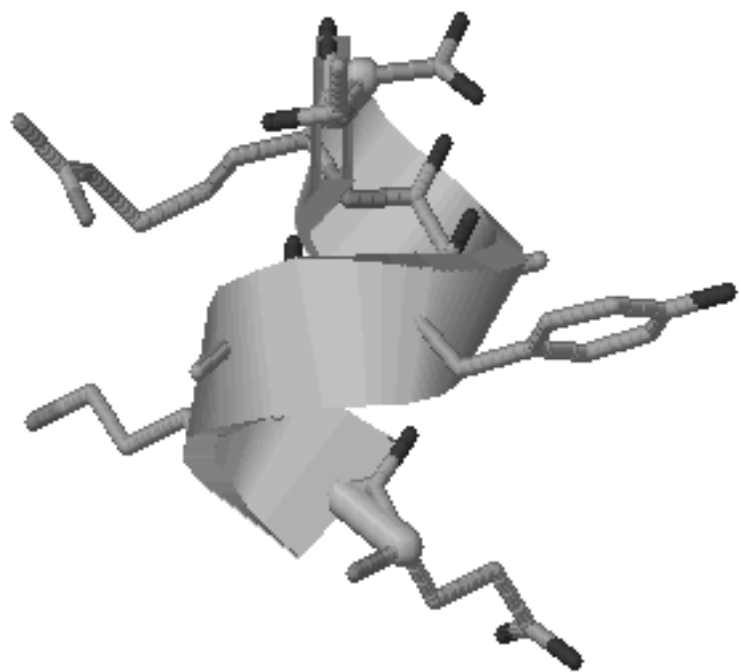
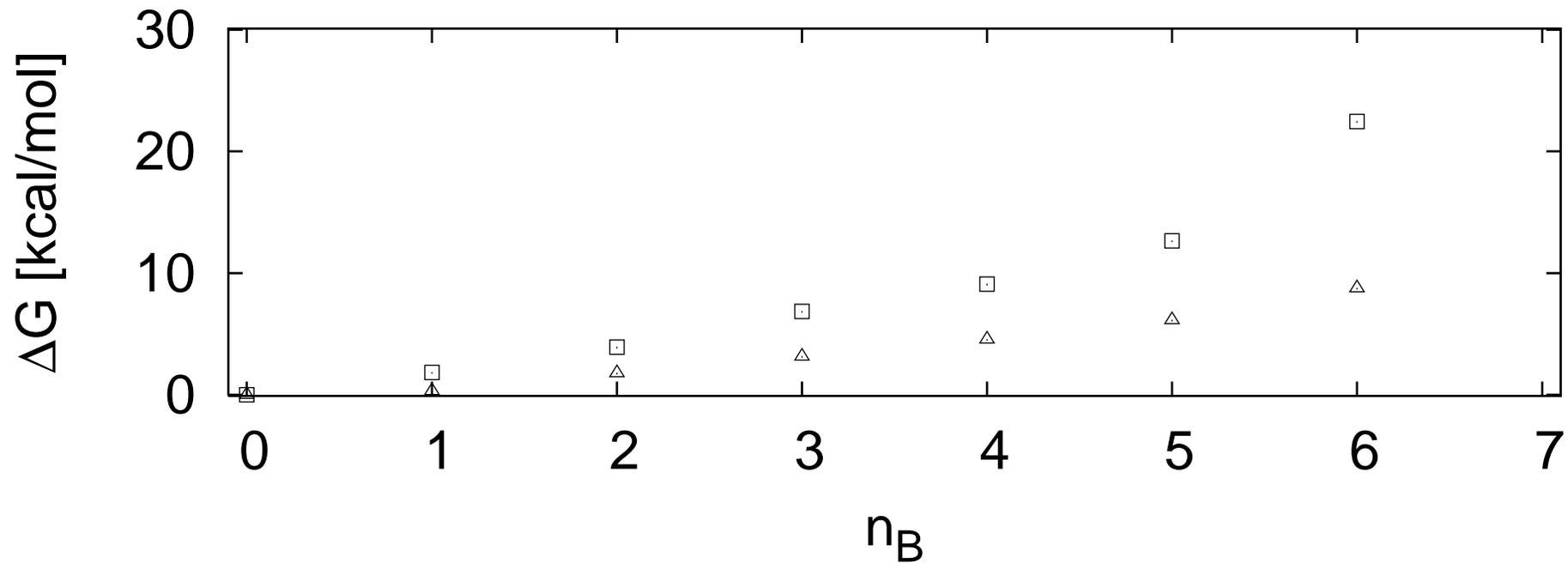
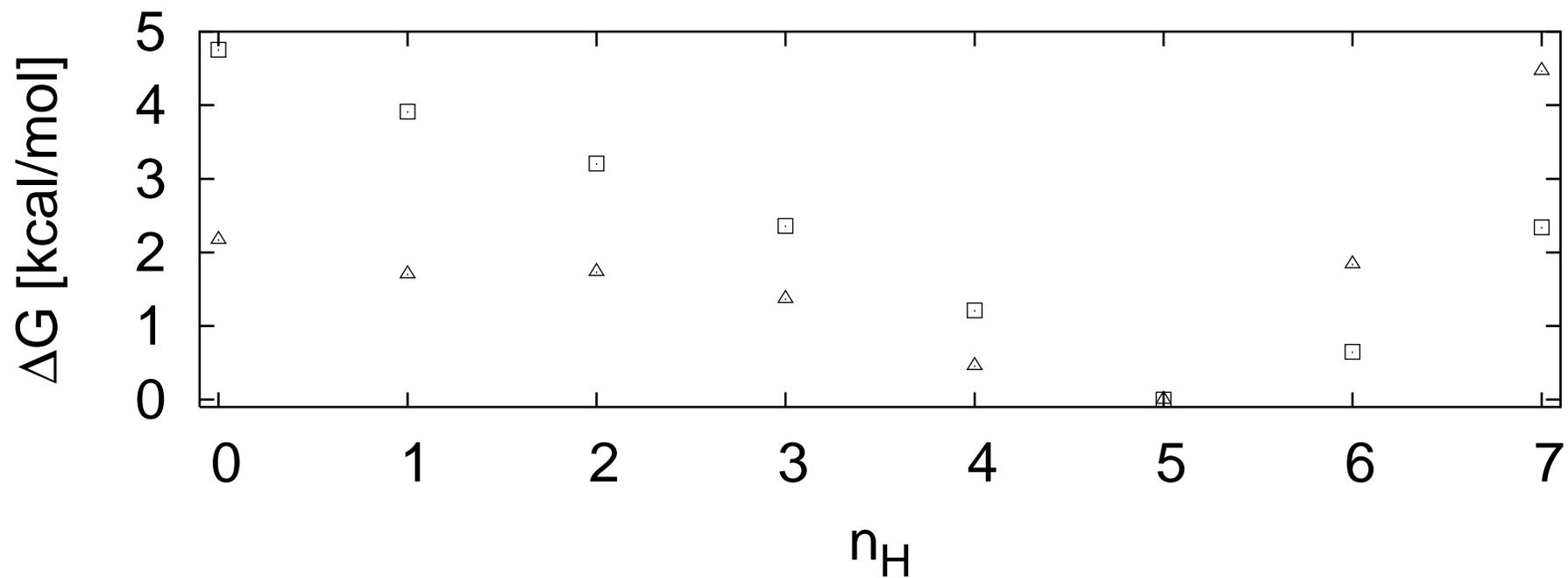
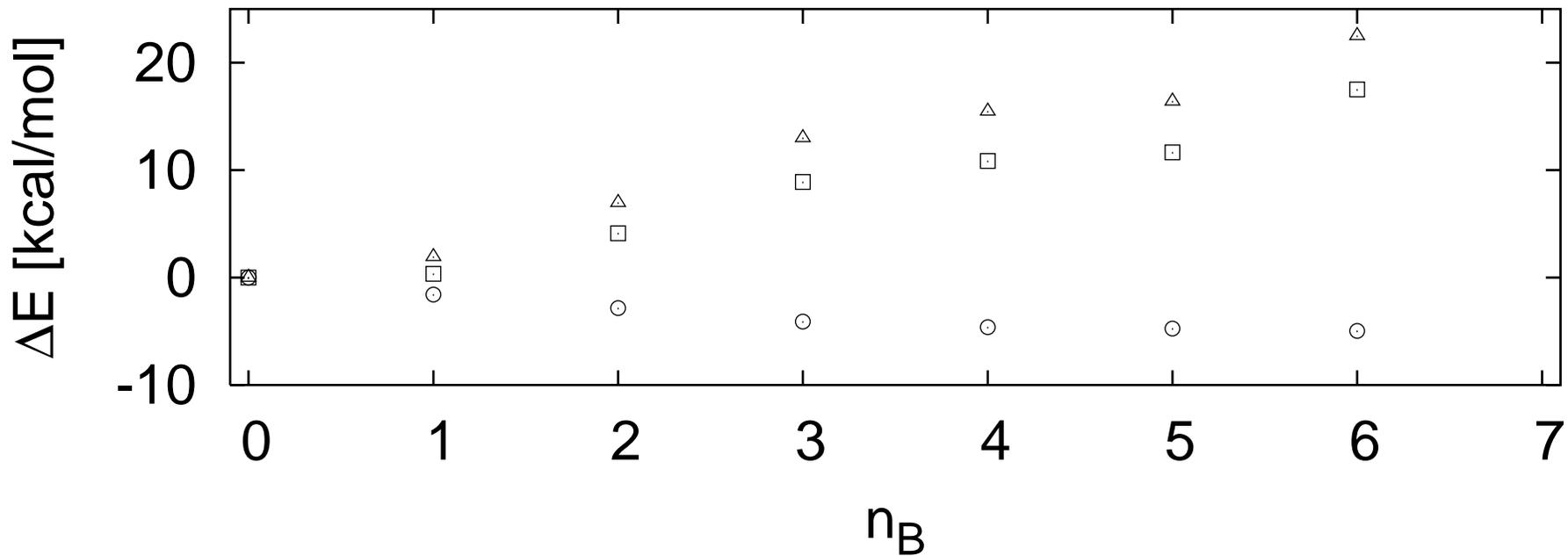
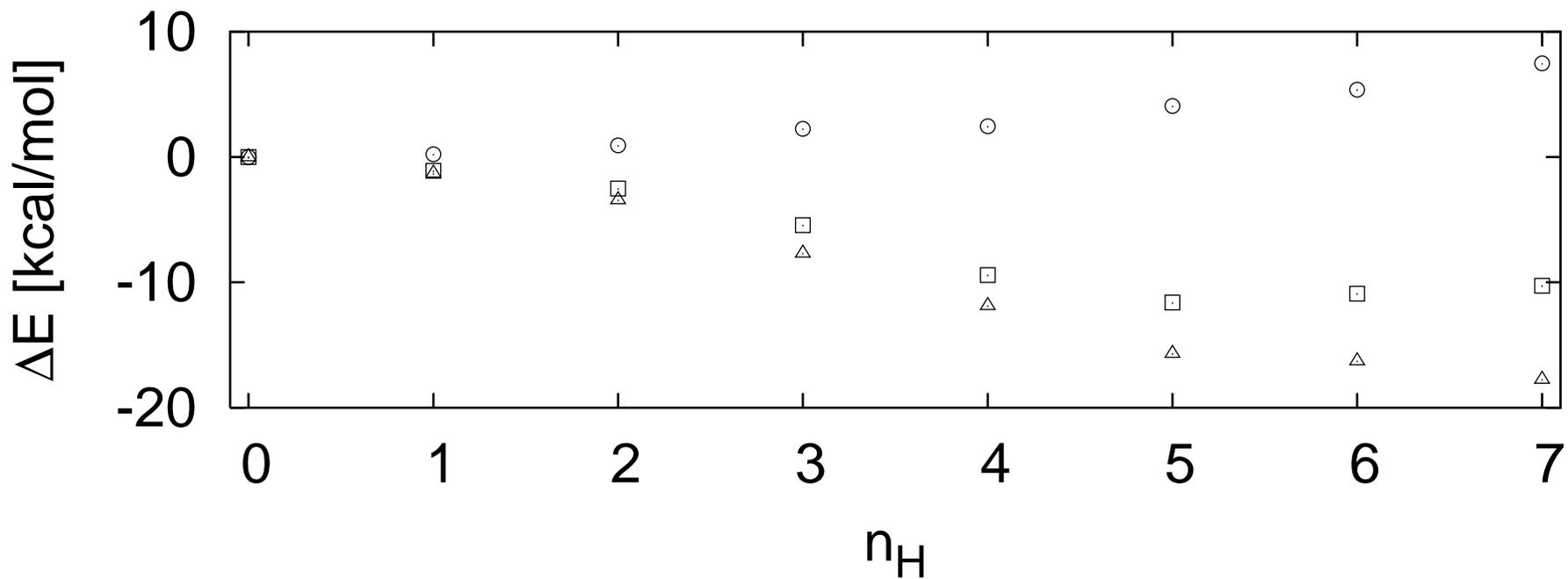
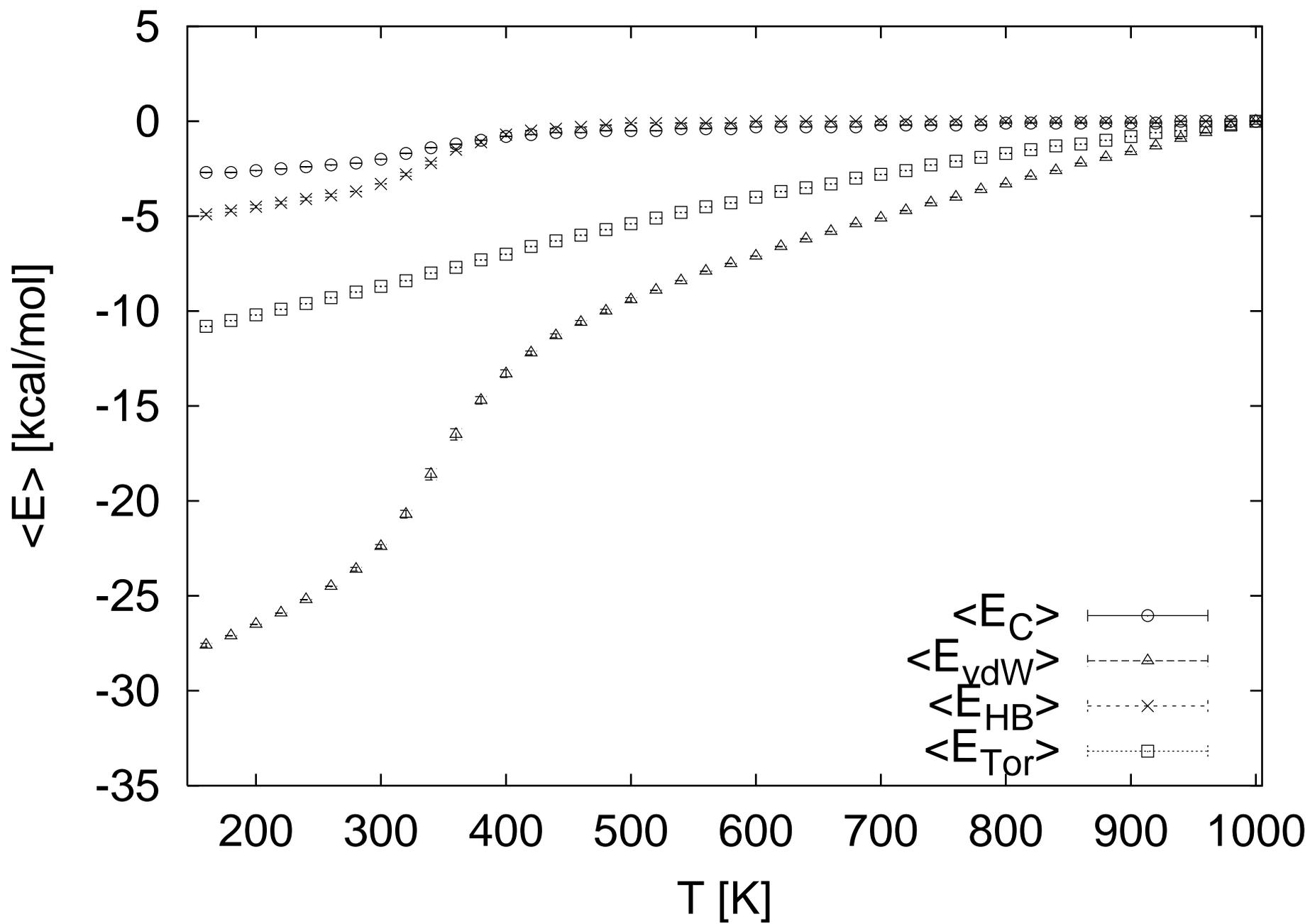
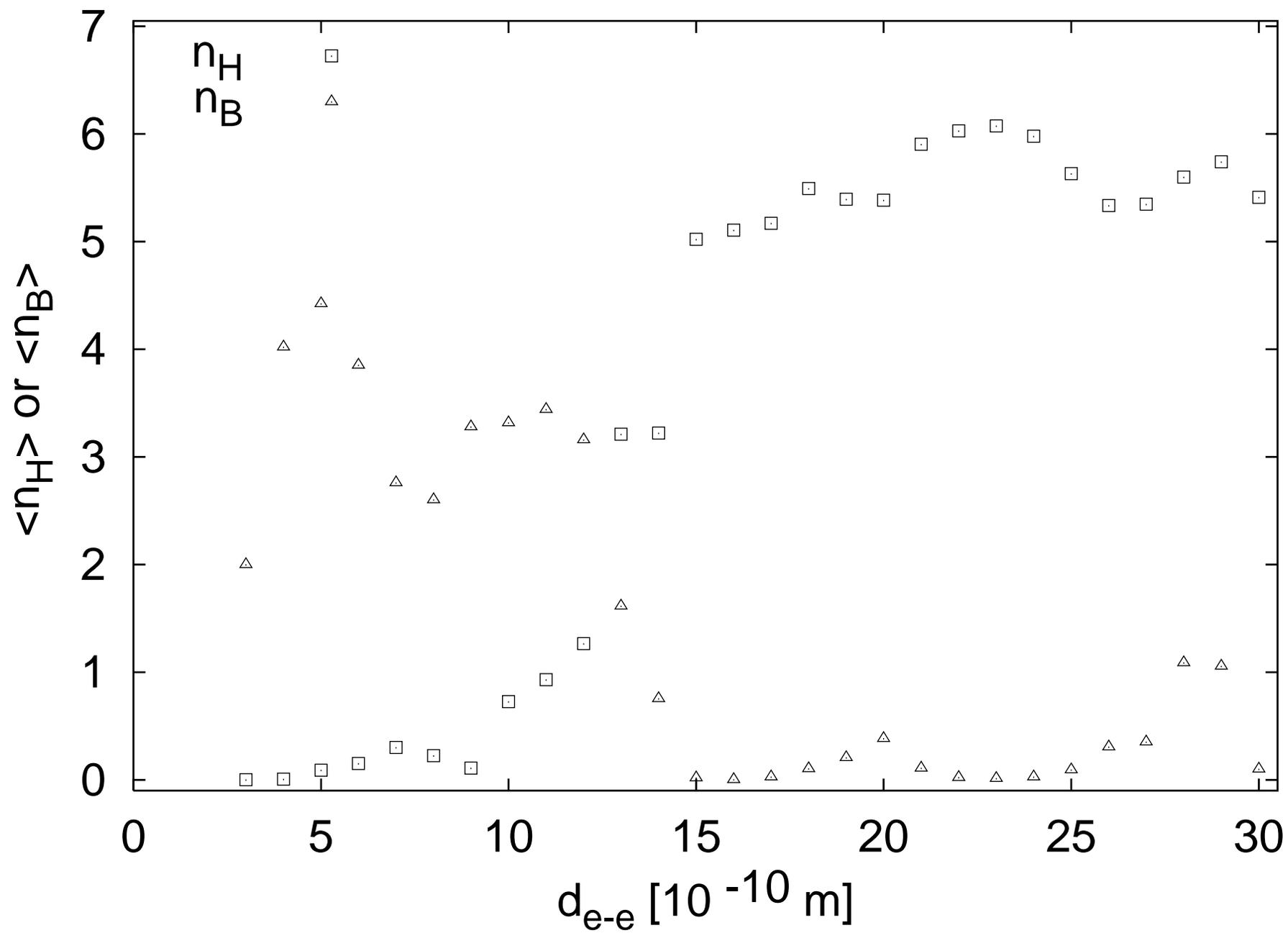
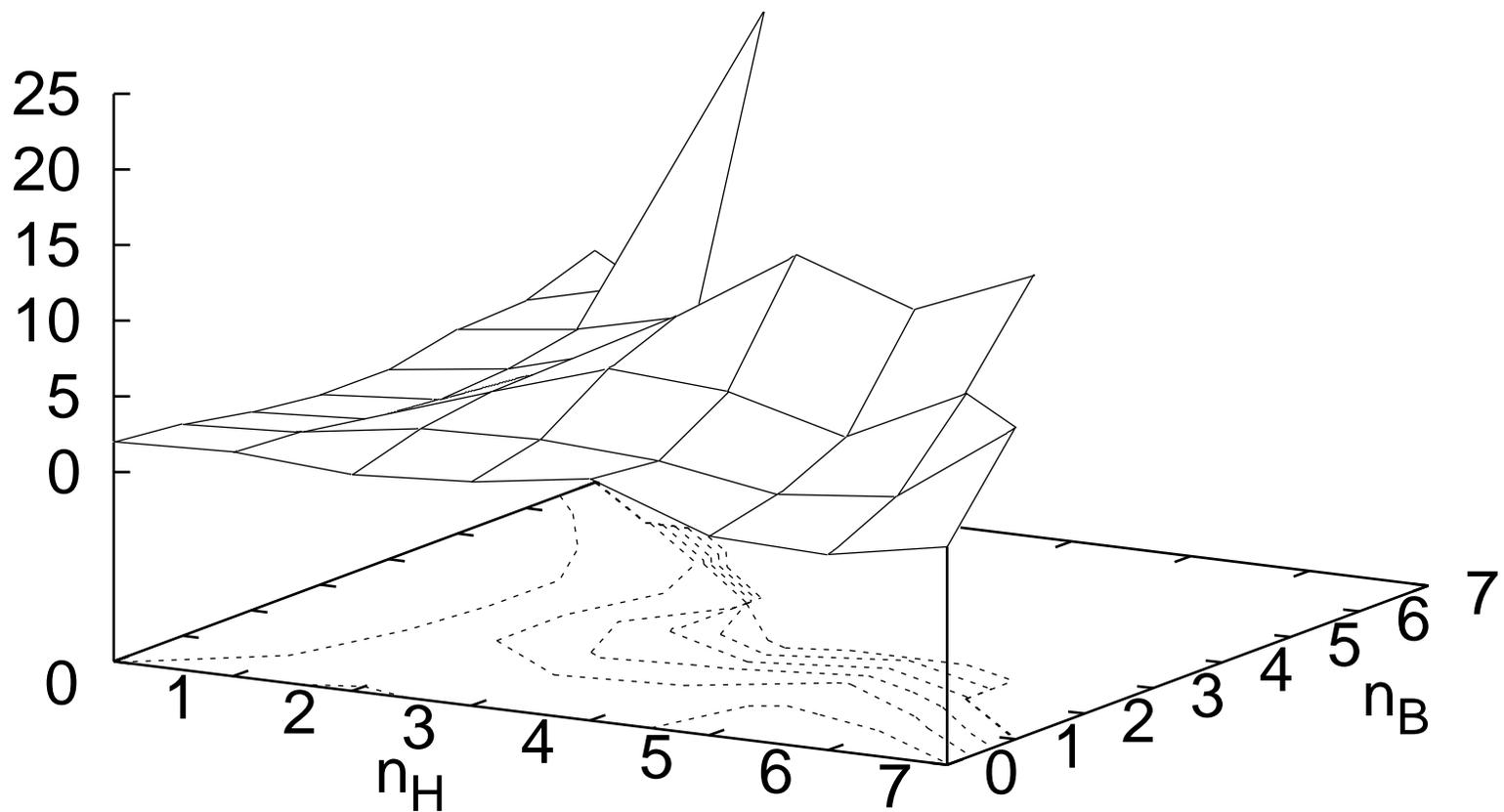
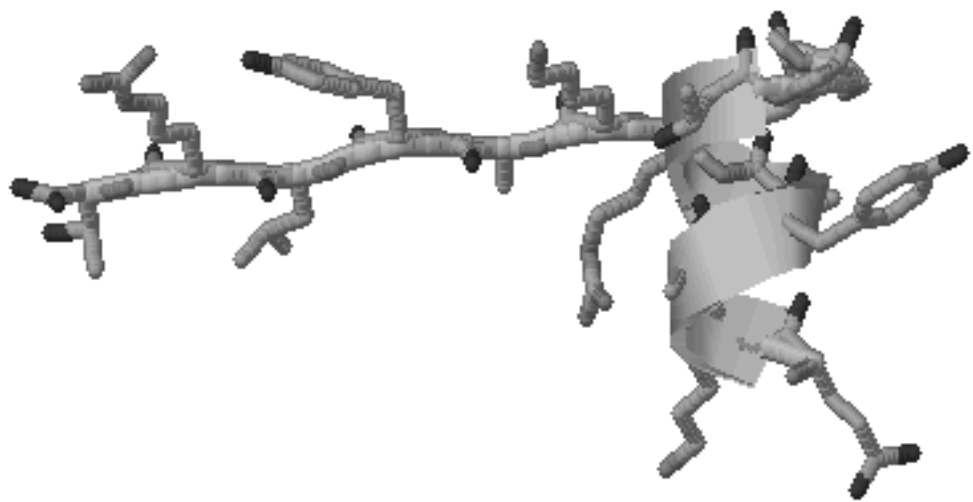
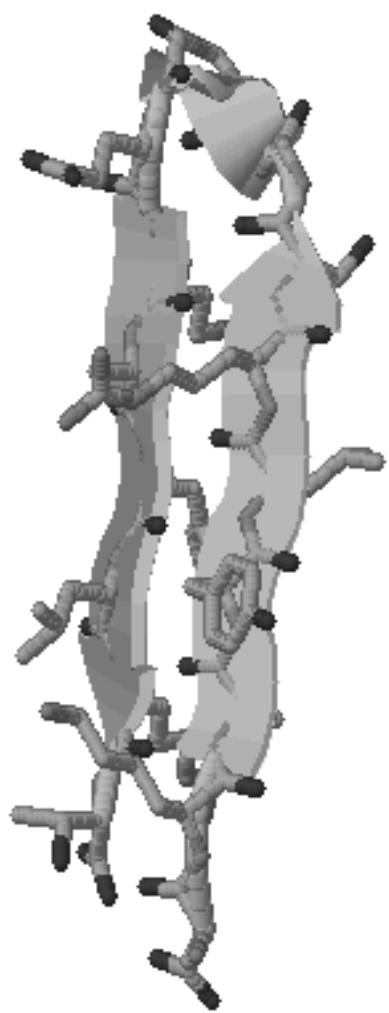
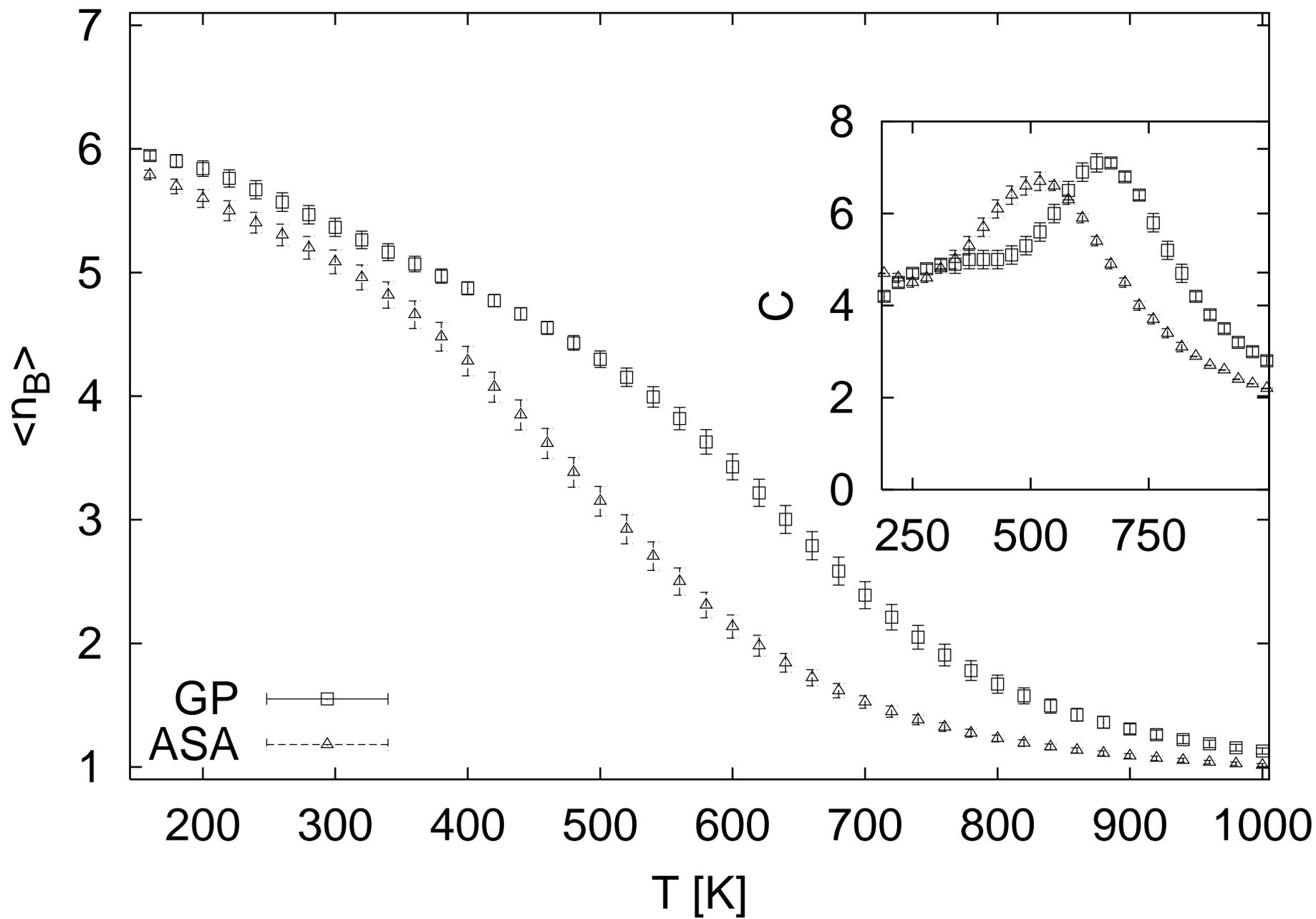
\end{document}